# Absolute phase measurement method based on bidirectional coding patterns

Xingyang Qi [a], Canlin Zhou [*a], Yixiao Wang [a], Shuchun Si [a], Hui Li [a]
a School of Physics, Shandong University, Jinan 250100, China
* canlinzhou@sdu.edu.cn; phone +86 13256153609

Abstract

The stair-phase-coding patterns have been widely used to determine the fringe order for phase unwrapping of the wrapped phase in 3D (three-dimensional) shape measurement. Although the special coding sequence algorithm can achieve with a large number of codewords, it needs current codeword and its adjacent codewords to jointly determine the fringe order. If any codeword of the grouped adjacent codewords is incorrectly recognized, it will result in false fringe order.Therefore, it is challenging to significantly increase the number of codewords. When it is necessary to simultaneously measure more than two isolated objects with large size differences, if the fringe frequency is high, the number of fringes on the smaller objects is too few to determine the fringe orders. On the other hand, if the fringe frequency is low, the image can't contain all isolated objects at the same time. To solve this problem, we propose an absolute phase measurement method based on bidirectional coding patterns. The wrapped phase of the object is obtained by four-step phase-shifting patterns, and the fringe orders is obtained by two bidirectional coded patterns. When coding the bidirectional patterns, we code two groups of stair-phase with different frequency along horizontal direction, which respectively represent local fringe order information and partition information, then, we alternately repeated the two groups of stair-phase along the vertical direction to obtain the bidirectional coding patterns in the whole pattern. Each local fringe order information and the corresponding partition information in small region jointly determine the fringe order of each position in the wrapped phase. To verify the effectiveness of our method, we did simulations and three experiments. Simulation and experimental results show that our method is effective for complex surfaces and isolated objects with different sizes.

1.Introduction

Fringe projection profilometry (FPP) has been widely applied to meet demands in diverse fields, including industrial production, quality control, reverse engineering, robot vision, and cultural relic protection, because of its advantages of fast measurement speed, low cost and high precision [1-8]. Because the arc-tangent function is introduced in the phase demodulation, the extracted phase is usually wrapped between $(-\pi, \pi]$. Therefore, many spatial and temporal phase retrieval methods have been presented for phase unwrapping [9-10]. Spatial methods include quality-guided, reliability-guided, least squares, and so on. These methods obtain the absolute phase by referring to the phase values of neighbor points, they can hardly measure an complex object or isolated parts. Temporal phase retrieval method can solve this problem. Popular TPU methods include two/multi-frequency method, phase coding (PC) method, gray code (GC) method, stereo-vision based method, deep learning, et.al. Among them, the two-frequency (TF) method minimizes the number of required patterns, and the hierarchical multi/two-frequency method shows the best unwrapping reliability, but it needs many frames of fringe patterns[11]. The phase coding (PC) method encodes the phase order into phase coding patterns [12]. The gray code (GC) method encodes the phase order into gray code patterns, but it can be easily influenced by the reflectivity of the tested object, ambient light and noise [13]. The stereo-vision based method projects a single random pattern to build coarse camera -projector correspondence for phase unwrapping. However, it requires an additional camera and complex calibration, it increases the cost and limits the measurement area [14-15]. The TPU based on Deep-learning uses only two (one unit-frequency, one high-frequency) wrapped phases as input, and directly outputs an unwrapped phase map [16-18]. It can substantially improve the unwrapping reliability, However, the design and selection of the network structure, the preparation of training data set, and the training and tuning of the network are not trivial.

Though the stair-phase-coding method is robust to surface contrast, ambient light and camera noise and is extensively used in 3D shape measurement, there are two limitations in this method. They are respectively how to obtain the large number of codewords and to correct misalignment between the fringe order and wrapped phase.

In phase-coding method, N quantized phases with a stair height of $2\pi/N$ can be determined from the coded phase in the range of (0, $2\pi$] [19], where each quantized phase represents a codeword. A large number of codewords is required for accurate 3-D shape measurement because fringe patterns with high-frequency can efficiently improve measurement accuracy [20-21]. However, the phase domain is limited because of the inevitable interference from image blurring, system nonlinearity, and noise. Owing to the phase errors introduced by the random noises, projector/camera defocusing and image blurring, the boundary of adjacent fringe periods cannot be well defined, resulting in jump phase errors, the fringe order may shift left or right by a few pixels in the region of the $2\pi$ phase jumps corresponding to wrapped phase, Thus the fringe order misaligns with the wrapped phase which results in fringe order errors. To solve these two limitations, many scholars have put forward many solutions. For example, Xing et al. [22] combined the coded phase with the system nonlinearity compensation, Zheng et al. [20] presented a method to increase the number of codewords by coding two phase information. Zhou et al. [19] proposed a method that embedded sinusoidal fringe and stair phase fringe patterns into different color channels to reduce the number of patterns. Chen et.al [21] proposed a two-digit phase-coding (TDPC) strategy, in which each fringe of the sinusoidal phase-shifting patterns corresponds to a two-digit codeword encoded in the phase-coding patterns to increase the number of codewords. Wang et.al [23] proposed an enhanced phase-coding method to address this misalignment problem by using half-period codewords, in which each codeword is aligned to the half-period of the sinusoidal patterns. Zheng et.al [24] studied a joint coding strategy to perform the coding of high-density stripes, in which, composite fringe pattern can achieve large-number codewords considering both the phase domain and the intensity domain. Zhang et.al. [25] proposed a robust principal component analysis (RPCA)-based approach to remove the impulsive fringe order errors. Wang et.al [26] proposed a complementary phase-coding (CPC) method to address the misalignment between the fringe order and wrapped phase. Wu et.al [27] proposed a generalized tripartite phase unwrapping (Tri-PU) method to avoid rather than correct jump errors. Cai et.al [28] proposed an optimized phase-coding method to overcome the misalignment problem, in which half-period mask and K-means algorithm are combined to correct the codewords errors. Lv et.al [29] proposed an optimized phase-coding demodulation method to increase the number of codewords, in which, fringe orders are uniquely determined by the four adjacent coding phases, Chen et.al [30] proposed an absolute phase retrieval method based on quantized phase modulation and connected region labeling, in which fringe order is determined by 3-digit-codes combining the current period and its neighbors, Tian et.al [31] proposed a phase coding method based on special coding sequence of stair phase to obtain more large-number codewords. Cai et.al [32] proposed an optimized shifting phase coding method to achieve with a large number of codewords, in which the fringe order is determined using unique three-adjacent-codes combining the current period and its neighbors. Li et.al [33] proposed fringe-width encoded fringe patterns method, in which the encoded fringe-width sequence based on De-Bruijn sequence is used to identify the fringe order.

Through careful analysis, we found that the existing stair-phase-coding method only encodes along a single direction in spatial domain (that is the direction of stripe change), especially in these references [29-33], authors adopted the special coding sequence along horizontal direction to achieve with a large number of codewords. In the pre-designed special coding sequence, every three (or many) adjacent-code is unique, each fringe period can be uniquely identified by looking up the position of these unique adjacent codewords in the whole code sequence. The more adjacent codewords that uniquely determine the fringe order, the greater the number of coded fringe orders. In general, adjacent codeword should be at least 3. Because multiple adjacent codewords over a larger range along the horizontal direction are required to determine the fringe order, if any codeword of the grouped adjacent codewords is incorrectly recognized, it will result in false fringe order. That is to say, although the special coding sequence method increases the number of codewords, it also increases the probability of mis-identification at the same time. Thus, in these references [29-33], the corresponding error correction algorithms are proposed and adopted. However, the process of error correction is not an easy thing.

Therefore, accurately measuring for objects with complex surfaces using the stair-phase-coding method remains a challenge.

We here aim to achieve with a large number of codewords and reduce the probability of mis-identification in stair-phase-coding method. In this paper, we proposed an absolute phase measurement method based on bidirectional joint coding strategy. We designed two types of stair-phase with different frequency in horizontal direction, the first group of stair-phase has frequency value same as the wrapped phase to record the local fringe order information, another group of stair-phase to record the partition information. In the vertical direction, the two groups of stair-phase are repeated alternately in the whole pattern. Therefore, the final fringe order of the wrapped phase can determine jointly using the local fringe order and partition information within a small region. The proposed method has the advantages of large number of codewords and low mis-identification rate for 3D surface measurement. Simulation and experimental results show that the proposed method is effective and accurate.

## 2.Principle

2.1 Four-step phase-shifting method

We obtain the wrapped phase of the object by the four-step phase-shifting method. The four-step phase-shifting fringe patterns can be written as:

$$I_1^P(x,y) = a(x,y) + b(x,y)\cos(2\pi f_0 x) \#(1)$$

$$I_2^P(x,y) = a(x,y) + b(x,y)\cos\left(2\pi f_0 x + 1 \cdot \frac{2\pi}{4}\right) \#(2)$$

$$I_3^P(x,y) = a(x,y) + b(x,y)\cos\left(2\pi f_0 x + 2 \cdot \frac{2\pi}{4}\right) \#(3)$$

$$I_4^P(x,y) = a(x,y) + b(x,y)\cos\left(2\pi f_0 x + 3 \cdot \frac{2\pi}{4}\right) \#(4)$$

where $(x,y)$ is the coordinate of the pixel, $a(x,y)$ and $b(x,y)$ denote the average intensity and intensity modulation, $f_0$ is the fringe frequency. We set both $a(x,y)$ and $b(x,y)$ to 0.5.

The fringe patterns captured by the camera can be described as:

$$I_1(x,y) = A(x,y) + B(x,y)\cos[\varphi(x,y)] \#(5)$$

$$I_2(x,y) = A(x,y) + B(x,y)\cos\left[\varphi(x,y) + 1 \cdot \frac{2\pi}{4}\right] \#(6)$$

$$I_3(x,y) = A(x,y) + B(x,y)\cos\left[\varphi(x,y) + 2 \cdot \frac{2\pi}{4}\right] \#(7)$$

$$I_4(x,y) = A(x,y) + B(x,y)\cos\left[\varphi(x,y) + 3 \cdot \frac{2\pi}{4}\right] \#(8)$$

where $A(x,y)$ and $B(x,y)$ denote the average intensity and intensity modulation, and $\varphi(x,y)$ denotes the phase of the tested object. After solving the equation (5)-(8), the variables $A(x,y)$、$B(x,y)$ and $\varphi(x,y)$ can be obtained as follow:

$$A(x,y) = \frac{[I_1(x,y) + I_2(x,y) + I_3(x,y) + I_4(x,y)]}{4} \#(9)$$

$$B(x,y) = \frac{\sqrt{[I_1(x,y) - I_3(x,y)]^2 + [I_4(x,y) - I_2(x,y)]^2}}{2} \#(10)$$

$$\varphi(x,y) = \arctan\frac{I_4(x,y) - I_2(x,y)}{I_1(x,y) - I_3(x,y)} \#(11)$$

Due to the arctangent operation, the phase of the object is wrapped between $[-\pi, +\pi)$. If we know the fringe order $k(x,y)$, we can calculate the absolute phase:

$$\Phi(x,y) = \varphi(x,y) + k(x,y) \times 2\pi \#(12)$$

Where $\varphi(x,y)$ is the wrapped phase, $\Phi(x,y)$ denotes the absolute phase.

2.2 Bidirectional coded stair phase patterns

In general, when higher fringe frequency is used to improve measurement accuracy, the number of fringe orders will be larger, which makes it difficult to decode the fringe orders. Some methods [29-33] encode special sequence of coding numbers and use several adjacent coding numbers to determine one fringe order. Because multiple coded numbers in a larger region along the horizontal direction are used together to determine one fringe order. This is not applicable when the number of fringes projected on the object is insufficient or lacking, and if any of the grouped adjacent codewords is incorrectly recognized, it will affect the judgment of a series of fringe orders.

Here, we proposed an absolute phase measurement method based on bidirectional coding patterns, which codes two types of stair-phase with different frequency in horizontal and alternately repeated coded stair-phase along vertical directions. We can use the stair-phase information to estimate the local fringe order and partition segment information, then determine the final fringe order. Our method is effective even if some fringes are missing, because they are confined to small local areas, do not affect the other regions.

We designed two types of stair-phase with different frequency in horizontal direction, the first group of stair-phase has frequency value *f*, which is same as the wrapped phase to record the local fringe order information, another group of stair-coding phase has frequency value *f/6* to record the partition information, In the vertical direction, the two sets of stair-phase are repeated alternately in the whole pattern. The detailed process is described below.

First, we need to determine how many regions to divide the bidirectional pattern into, and the number of stair phase in each region. The number of partitions times the number of stair phase in each region equals the total number of fringes. Here, for the convenience of explanation, we set the parameters as follows. We set the fringe frequency to 1/30 pixel, the number of partitions to 6, the number of stairs in each region to 6, the total fringe number to 36 and the resolution of patterns to 1080*1080 as examples.

Then, we encode local fringe order information and partition information horizontally according to the following equation:

$$\varphi_{s1}(x,y) = -\pi + \frac{1}{6}\cdot\pi + \frac{2}{6}\cdot\pi\cdot mod(floor\{x\cdot f\},6) \#(13)$$

$$\varphi_{s2}(x,y) = -\pi + \frac{1}{6}\cdot\pi + \frac{2}{6}\cdot\pi\cdot floor\left\{x\cdot\frac{f}{6}\right\} \#(14)$$

where $floor\{\}$ is rounding down function, $mod()$ is the remainder function, $\varphi_{s1}(x,y)$ denotes local fringe order information, this group of stair-phase has frequency value *f*, which is same as the wrapped phase, $\varphi_{s2}(x,y)$ denotes partition information, this group of stair-phase has frequency value *f/6*.

Fig. 1 (1) shows a row of $\varphi_{s1}$. Fig. 1 (2) shows a row of $\varphi_{s2}$.

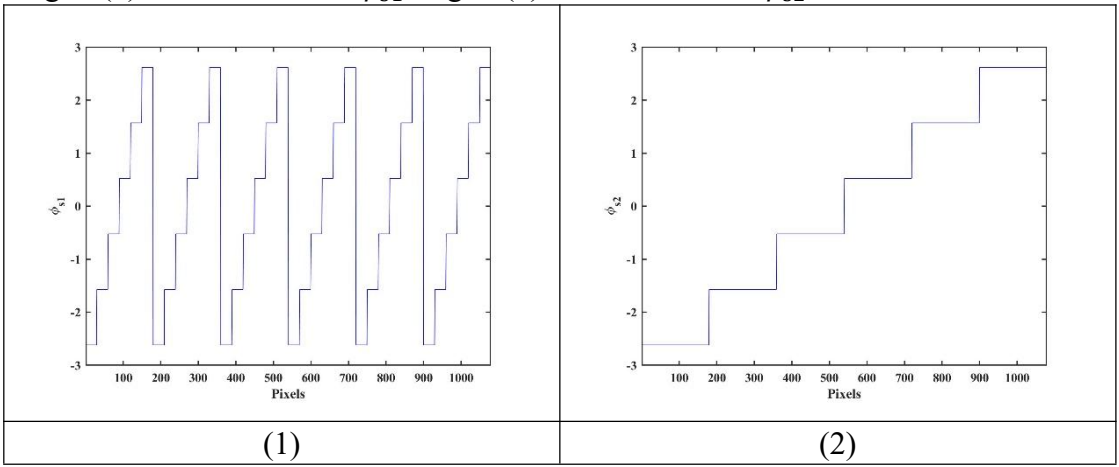

(1)　　　　　　　　　　　　　　(2)

Figure 1. Two types of stair-phase. (1) A row of $\varphi_{s1}$; (2) A row of $\varphi_{s2}$.

Then we alternate sine and cosine processed $\varphi_{s1}$ and $\varphi_{s2}$ in the vertical direction to form the bidirectional coded stair-phase patterns. Since local fringe order information and partition information will be identified later, the number of rows of local fringe order information and partition information must be different. We set the ratio of the number of rows of consecutive local fringe order information and partition information to 2:1. In the vertical direction of the bidirectional coded stair-phase pattern, each 30 rows of pixels are a large group. The first 20 rows encode local fringe order information (the first group), and the last 10 rows encode partition information (the second group). The above coding process is repeated until the entire bidirectional coded stair phase pattern is formed. The formula is as follows:

$$I_5^P(x,y) = \begin{cases} a(x,y) + b(x,y)\cos(\varphi_{s1}(x,y)), & mod(y,30) \leq 20 \\ a(x,y) + b(x,y)\cos(\varphi_{s2}(x,y)), & mod(y,30) > 20 \text{ or } mod(y,30) = 0 \end{cases} \#(15)$$

$$I_6^P(x,y) = \begin{cases} a(x,y) + b(x,y)\sin(\varphi_{s1}(x,y)), & mod(y,30) \leq 20 \\ a(x,y) + b(x,y)\sin(\varphi_{s2}(x,y)), & mod(y,30) > 20 \text{ or } mod(y,30) = 0 \end{cases} \#(16)$$

where $I_5^P$ denotes the bidirectional coded stair phase pattern, $I_6^P$ denotes the bidirectional coded stair phase pattern after π/2 phase shift.

Fig. 2 is a schematic diagram of bidirectional coded stair phase pattern.

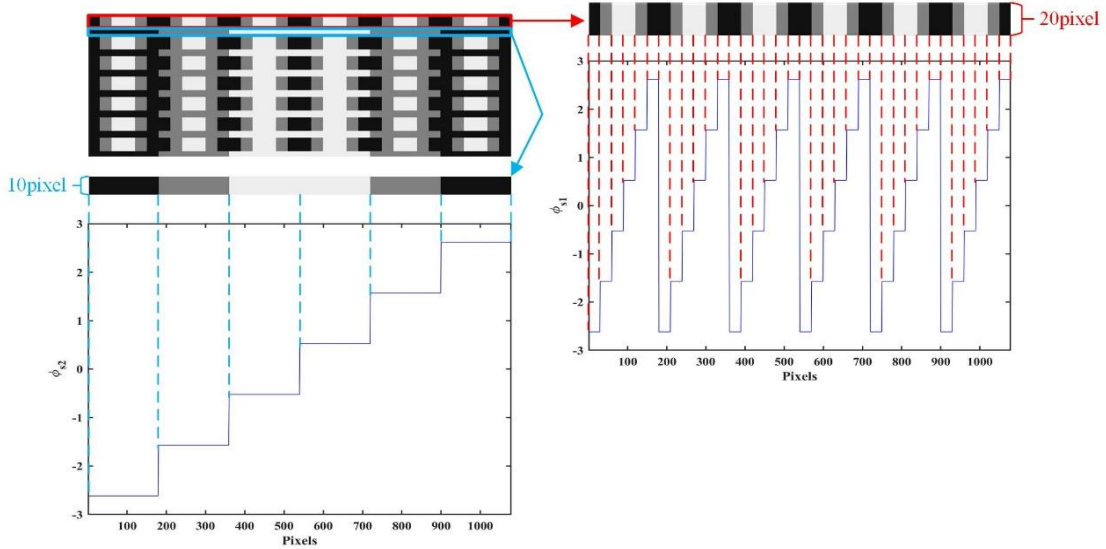

Figure 2. Schematic diagram of bidirectional coded stair phase pattern.

The part circled by red line in the upper left part of Fig. 2 is 20 rows of local fringe information. The part circled by blue line in the upper left part of Fig. 2 is 10 rows of partition information. These 30 rows are one large group. We repeat this encoding along the vertical direction, and we will obtain the final pattern.

Fig. 3 shows the bidirectional coded stair phase patterns. Fig. 3(1) is the bidirectional coded stair phase pattern and Fig. 3(2) is the bidirectional coded stair phase pattern after π/2 phase shift.

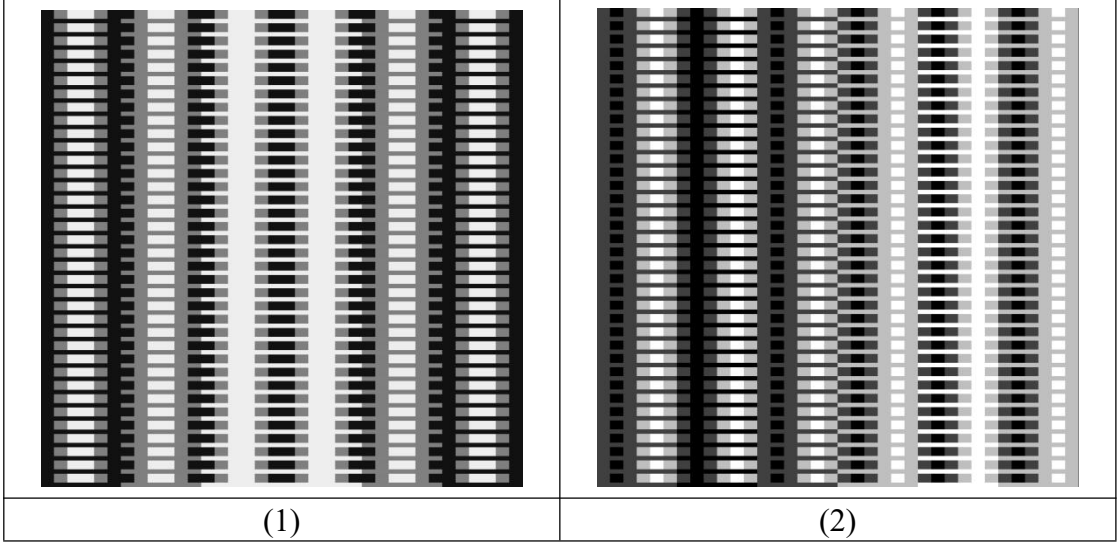

Figure 3. The bidirectional coded stair phase patterns. (1) The bidirectional coded stair phase pattern; (2) The bidirectional coded stair phase pattern after π/2 phase shift.

In addition, the ratio of the number of rows of adjacent local fringe order information and partition information, and the number of rows in each group can be set to other values. Through experiments, we found that it is easier to separate the local fringe order information and partition information when the ratio of the number of them above 2:1. And the fewer rows in each group, the fewer pixels used to determine the fringe order of this group, and the lower the impact of errors. However, the number of rows should not be too small, otherwise the local fringe order information and partition information will be difficult to identify.

2.3 The decoding principle
Project two bidirectional coding patterns on the object, and take the deformed bidirectional coding patterns modulated by the object. Then we can determine the fringe order k by decoding captured bidirectional coding patterns.
Project two bidirectional coded stair phase patterns $I_5^P$ and $I_6^P$ onto the object, and the stair phase patterns taken by the camera can be written as:
$$I_5(x,y) = A(x,y) + B(x,y)\cos[\varphi_s(x,y)] \quad \#(17)$$
$$I_6(x,y) = A(x,y) + B(x,y)\sin[\varphi_s(x,y)] \quad \#(18)$$
where $\varphi_s(x,y)$ denotes stair phase.
Combining the Eq. (9) obtained above and Eq. (17)-(18), we can subtract $A(x,y)$ and solve $\varphi_s(x,y)$:

$$\varphi_s(x,y) = \arctan\left(\frac{I_6(x,y) - A(x,y)}{I_5(x,y) - A(x,y)}\right) \#(19)$$

Then, we solve the compound local fringe order information and partition information: $k_s(x,y) = round\left\{\left[\varphi_s(x,y) + \frac{5}{6}\cdot\pi\right]\cdot\frac{6}{2\pi}\right\} \#(20)$

where $k_s(x,y)$ is compound fringe information and partition information, $round\{\}$

is the rounding function to get the nearest integer value.

Fig. 4 is the $k_s(x,y)$ of a plane as the object to be measured.

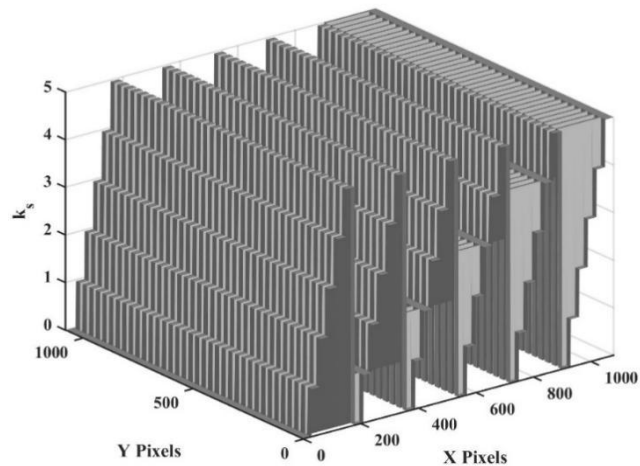

Figure 4. The $k_s$ of a plane as the object to be measured.

To get the final fringe order $k(x,y)$, we must separate the local fringe order information and partition information form the compound information. Since the number of rows of the local fringe order information is twice as many as the partition information, we can use this character to separate them.

Scan $k_s$ from top to bottom column by column, and the number of all adjacent pixels with the same $k_s$ value is recorded as C. The C value of local fringe order information is large, while the C value of partition information is small.

Each adjacent local fringe order information and partition information is defined as a group. Do the differential operation for C along the vertical direction. The boundary pixel of each group is the pixel whose difference of C is greater than 0, so that each column can be divided into many groups.

Then the fringe order of all pixels in one group can be obtained from the local fringe order information and partition information of this group. The final continuous fringe order $k(x,y)$ can be precisely calculated according to the following formula:

$$k(x,y) = k_{s1} + 6 \cdot k_{s2} \#(22)$$

where $k_{s1}$ denotes the local fringe information of one group, $k_{s2}$ denotes the partition information of the same group, $k(x,y)$ denotes the final continuous fringe order.

Fig. 5 shows how to calculate the fringe order $k$ from $k_s$ with an example of a column.

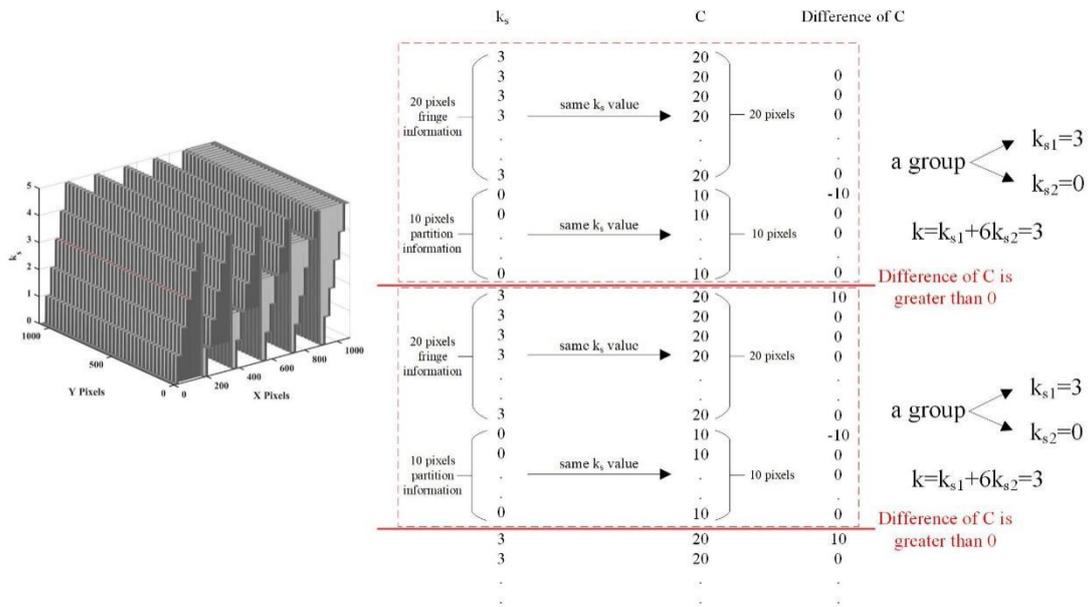

Figure 5. Fringe order calculation process.

In addition, when $k_{s1}$ and $k_{s2}$ are the same, the value of C is very large, and the fringe order can be directly calculated by 7 times $k_{s1}$.

Finally, the absolute phase can be restored by combining the wrapped phase with the fringe order. The calculation method is shown in Eq. (12).

Fig. 6 is a flow chart for our method.

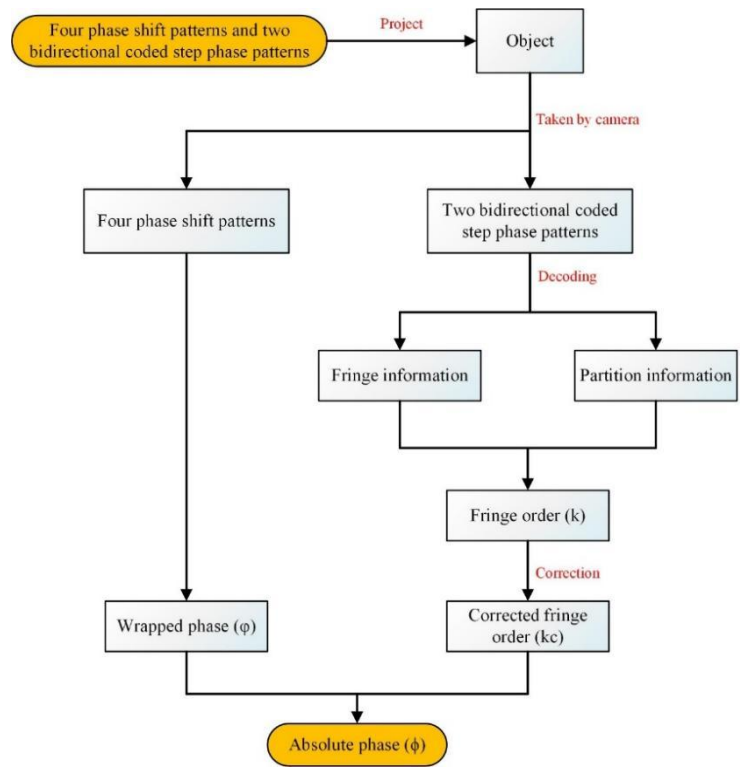

Figure 6. Flow chart for our method.

## 3. Simulation

To prove the superiority of our method, we use our method to simulate the measurement of one big object and three small objects in Matlab, and we compare the results with Lv's method.

First, we use our method to encode four-step phase-shifting patterns and two bidirectional coded stair phase patterns according to Eq. (1)-(4) and Eq. (13)-(16). The resolution is 1080*1080, the fringe frequency is 1/30 pixels. The number of partitions is 6, the number of stairs in each region is 6, the total fringe number is 36.

After that, we use Lv's method to encode four-step phase-shifting patterns and two stair phase patterns.

We add white Gaussian noise to simulated images. The mean is 0, and the variance is 1/10.

Fig. 7 shows the tested objects.

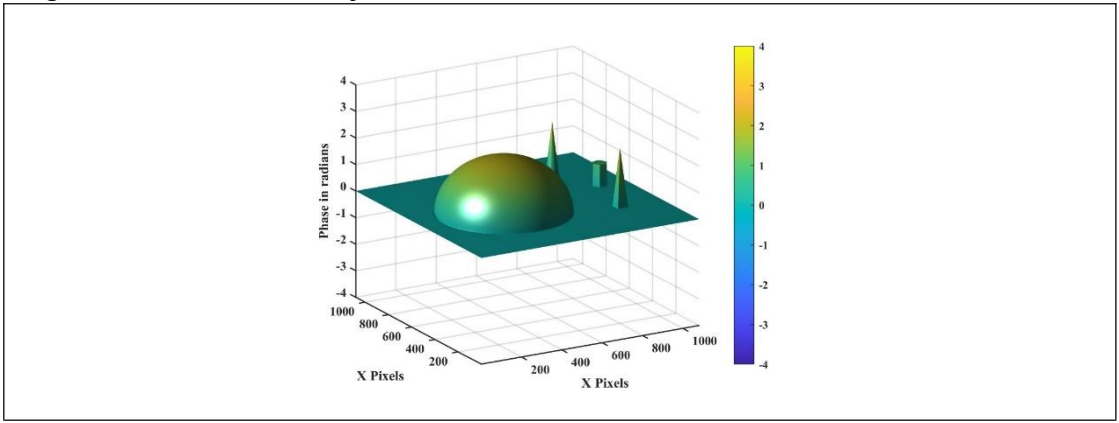

Figure 7. The tested objects.

Fig. 8 shows the deformed images of our method and Lv's method. Fig. 8(1)-(4) are the four-step phase-shifting patterns of our method, and Fig. 8(5)-(6) is the bidirectional coded stair phase patterns of our method. Fig. 8(7)-(12) are the patterns of Lv's method.

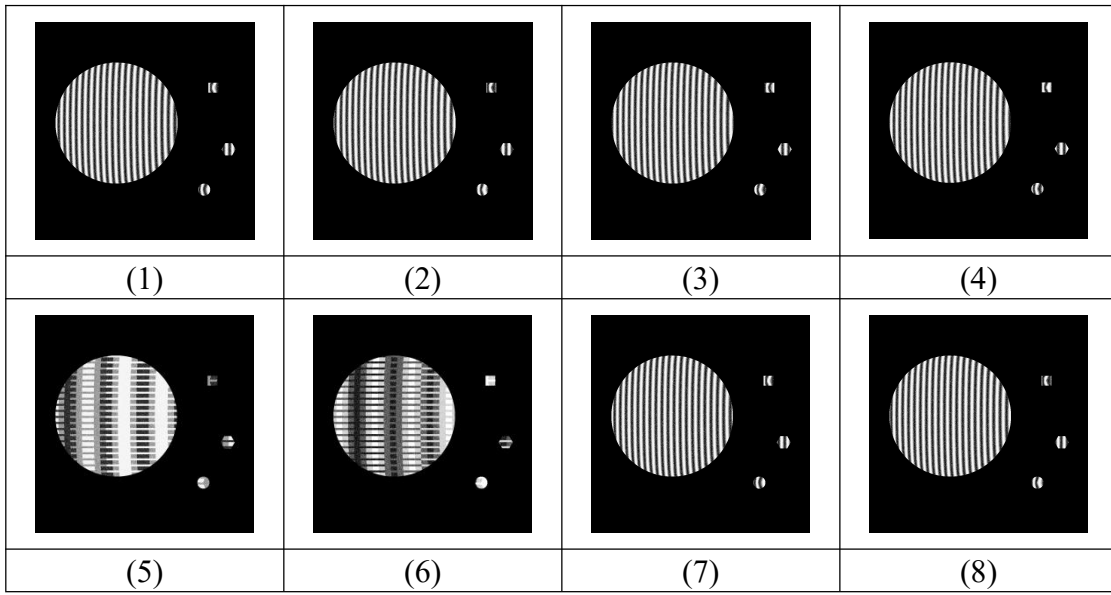

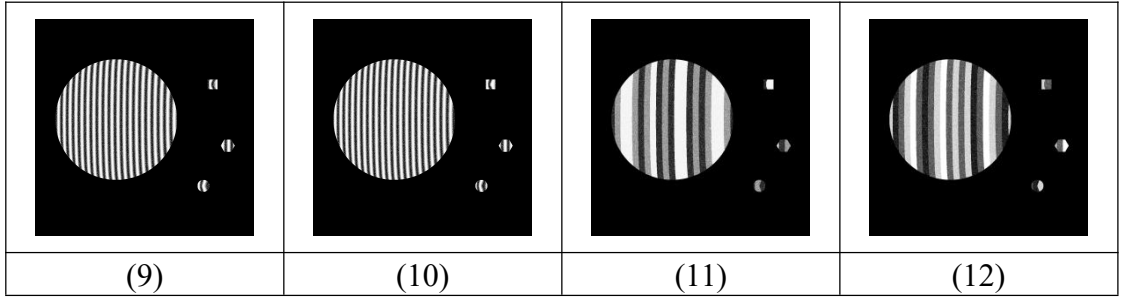

| (9) | (10) | (11) | (12) |

Figure 8. The deformed images of our method and Lv's method. (1)-(4) The four-step phase-shifting patterns of our method; (5)-(6) The bidirectional coded stair phase patterns of our method; (7)-(12) are the patterns of Lv's method.

Then, we process the deformed images. Here we describe the procedure of our method in detail.

We use four phase-shifting images to obtain the wrapped phase by Eq. (11) and average intensity by Eq. (9). We use bidirectional coded stair phase patterns get the $k_s(x, y)$ by Eq. (19)-(20). Then we scan $k_s$ column by column to distinguish fringe information and partition information. The fringe orders are calculated by Eq. (23). Because of the fringe distortion, some of the local fringe information and partition information are incorrectly identified, resulting in some incorrect fringe orders. So, we change the fringe orders of all pixels in one fringe to the mode of fringe orders in this fringe. Finally, we can obtain the absolute phase by Eq. (12).

Fig. 9 shows the simulation results. Fig. 9(1) is the wrapped phase. Fig. 9(2) is $k_s$. Fig. 9(3) is the fringe order.

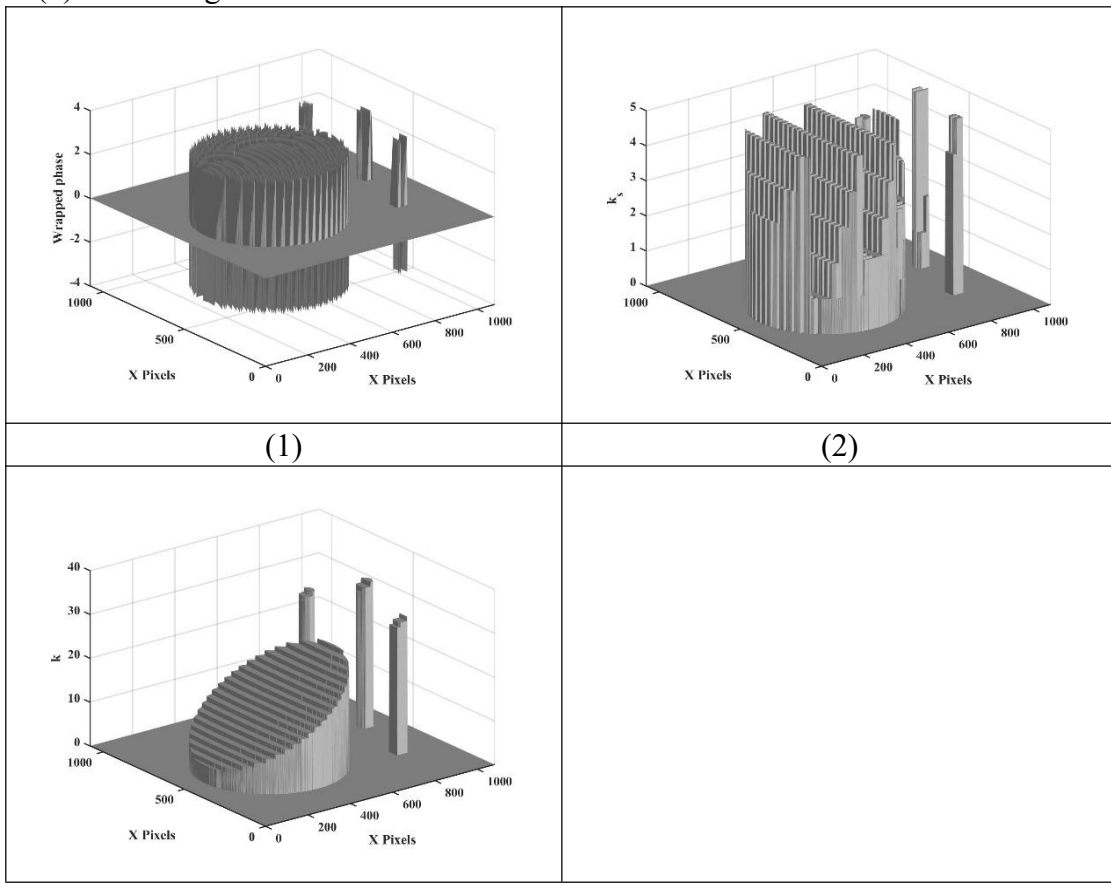

|  | (3) |
|---|---|

Figure 9. The simulation results. (1) The wrapped phase; (2) $k_s$; (3) The fringe order.

We compare the absolute phase recovered by our method and the absolute phase recovered by Lv's method. Fig. 10 shows the absolute phase. Fig. 10(1) is the absolute phase recovered by our method. Fig. 10(2) is the absolute phase recovered by Lv's method.

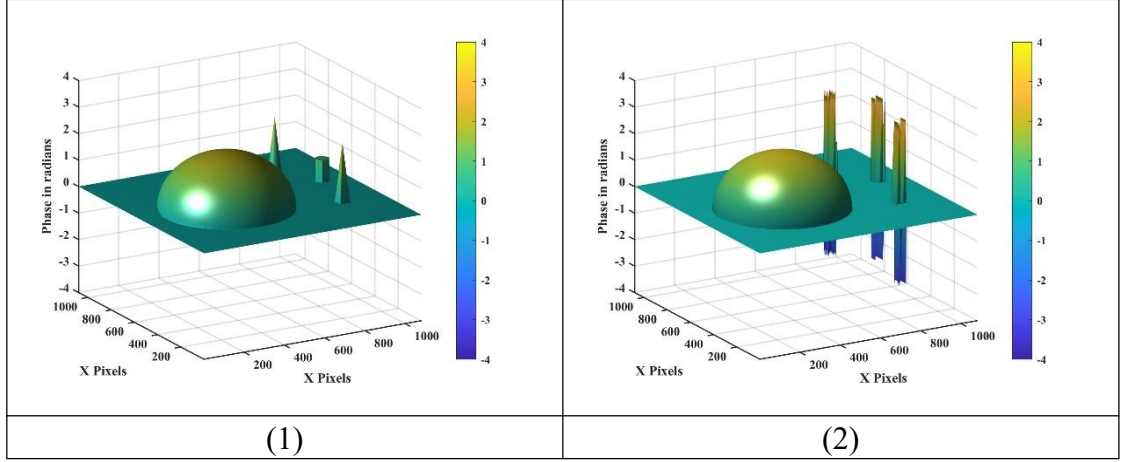

| (1) | (2) |
|---|---|

Figure 10. The absolute phase. (1) The absolute phase recovered by our method; (2) The absolute phase recovered by Lv's method.

It can be seen from Fig. 10 that we recovered the absolute phase of all objects, while the Lv's method only recovered the absolute phase of large objects due to the insufficient number of fringe orders on small objects. Our method only needs a set of local fringe order information and partition information in a very small range to determine the fringe order, but does not need the information of adjacent fringes. Our method has some advantages when dealing with objects with large size differences.

## 4. Experiment

To verify the performance of our method, we construct a digital fringe projection measurement system. The projector is Optoma EH412, and the resolution is 1920*1080. The camera is a Daheng digital camera (MER-130-30UM), and the resolution is 1280×1024. The camera sensor is 1/1.8'' Rolling Shutter Onsemi MT9M001 CMOS. The measurement software is programmed using MATLAB.

We designed three experiments, and the measured objects were two masks, one mask and one prism, two isolated sculptures.

In the first experiment, the tested objects are two isolated masks. They are measured by our method, Lv's method and three-frequency twelve-step phase-shifting method respectively. In our method, the fringe frequency is 1/30 pixel, the number of partitions is 6, the number of fringes in each region is 6, the total fringe number is 36. The fringe frequency and total fringe number of Lv's method are the same as ours. The fringe frequency of three-frequency twelve-step phase-shifting method is 1/1080 pixel, 1/180 pixel and 1/30 pixel respectively.

Fig. 11 are patterns of our method. Fig. 11(1) - (4) are four-step phase-shifting fringe patterns. Fig. 11(5) - (6) are bidirectional stair phase coding patterns.

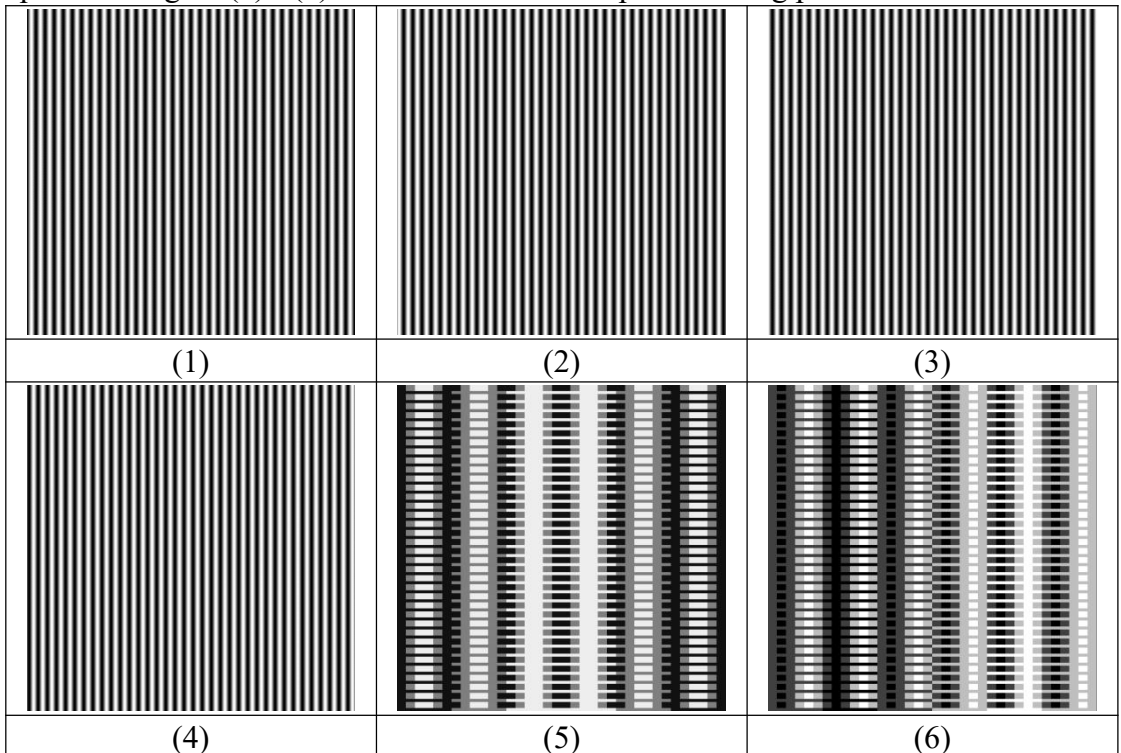

Figure 11. Patterns of our method. (1)-(4) Four-step phase-shifting fringe patterns. (5)-(6) Bidirectional coded stair phase coding patterns.

Fig. 12 are images taken by our method.

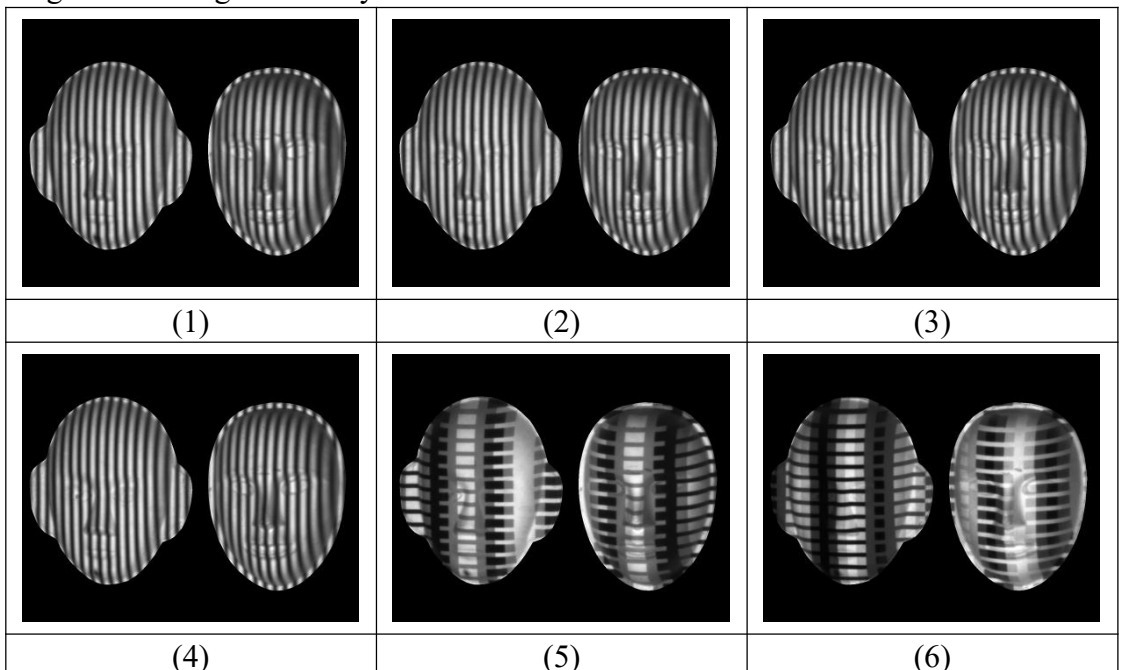

Figure 12. Images taken by our method. (1)-(4) Four-step phase-shifting fringe patterns. (5)-(6) Bidirectional coded stair phase coding patterns.

Fig. 13 are images taken by Lv's method.

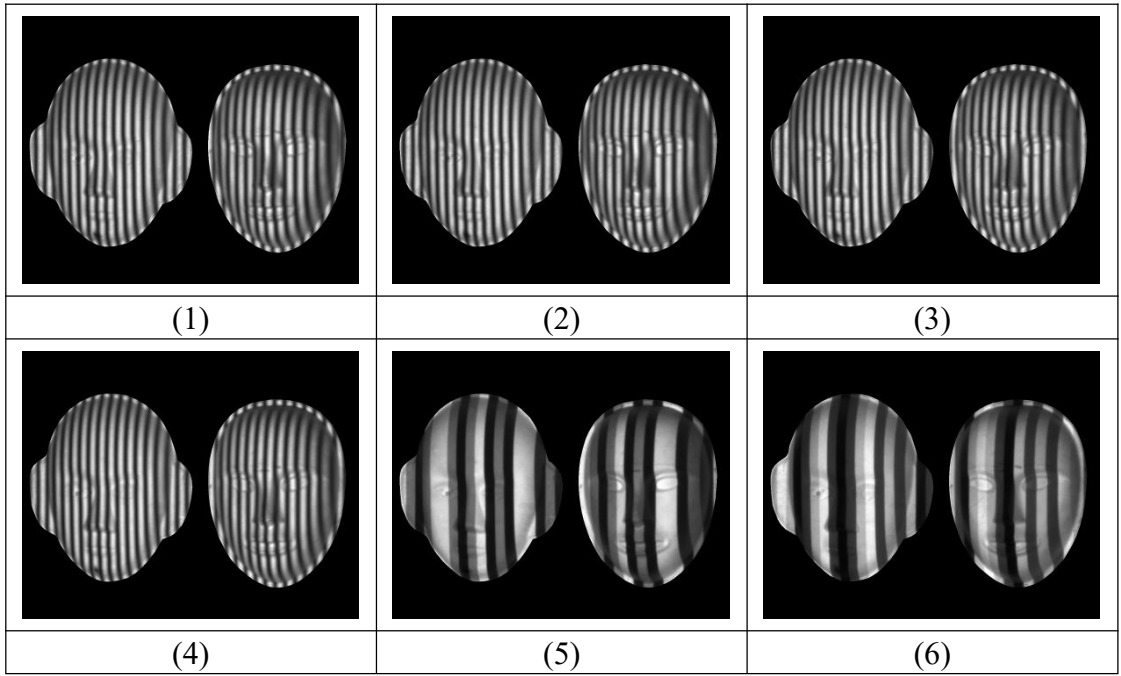

Figure 13. Images taken by Lv's method.

Fig. 14 are three of the thirty-six images taken by three-frequency twelve-step phase-shifting method images.

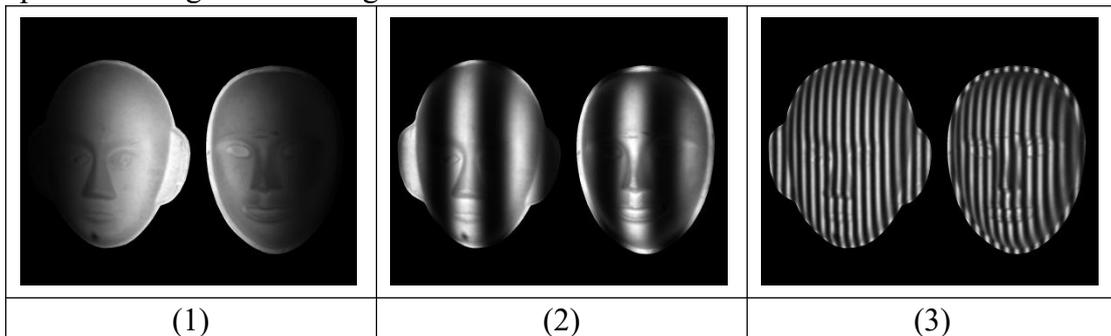

Figure 14. Three of the three-frequency twelve-step phase-shifting method images. (1) Frequency is 1/30 pixel; (2) Frequency is 1/180 pixel; (3) Frequency is 1/1080 pixel.

Fig. 15 is the processing result of our method. Fig. 10(1) is the local fringe order information and partition information, Fig. 10(2) is the fringe order.

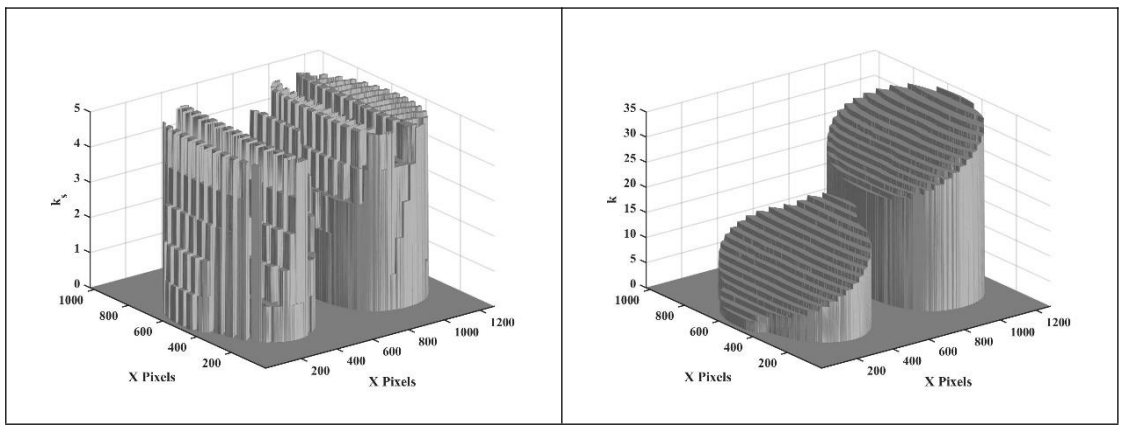

| (1) | (2) |

Figure 15. The processing result of our method. (1) The local fringe order information and partition information; (2) The fringe order.

Fig. 16 shows the absolute phase recovered by three methods. Fig. 16(1) shows the absolute phase recovered by our method. Fig. 16(2) shows the absolute phase recovered by Lv's method. Fig. 16(3) shows the absolute phase recovered by the three-frequency twelve-step phase-shifting method.

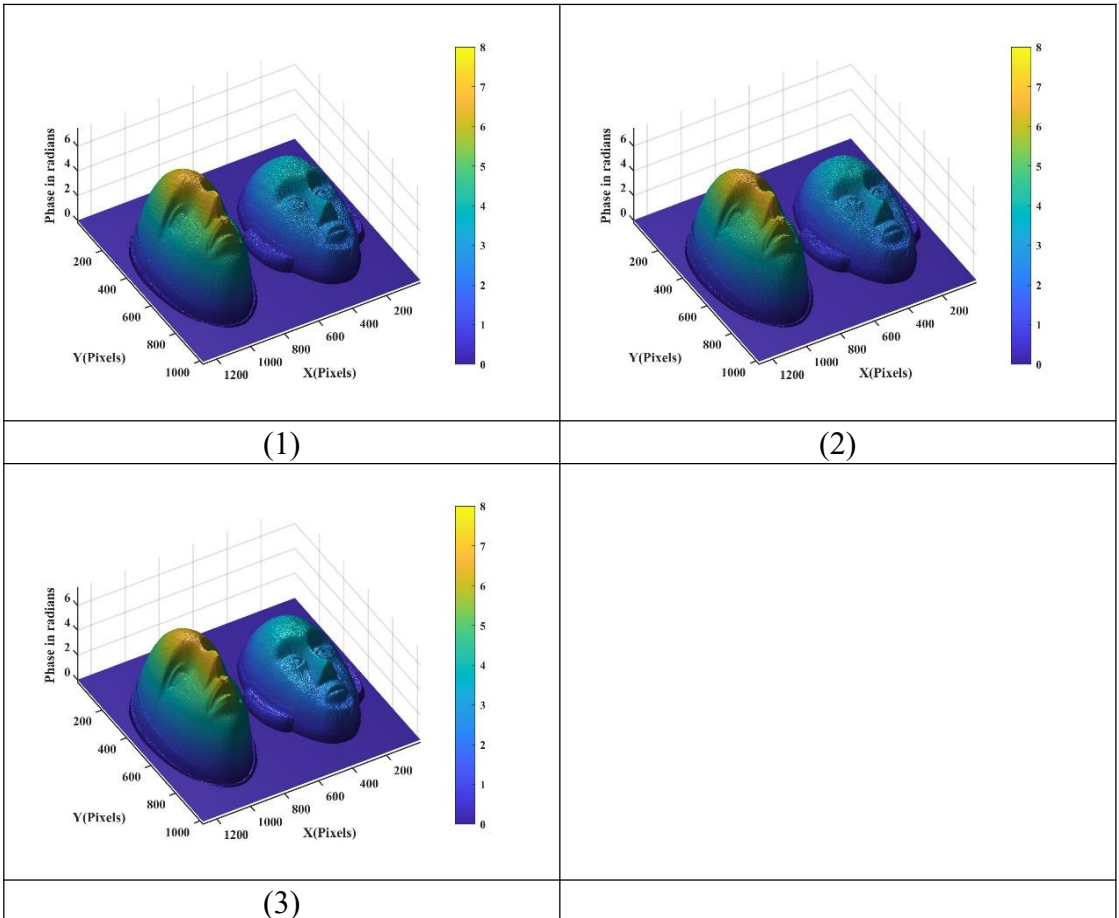

| (1) | (2) |
| (3) | |

Figure 16. The absolute phase recovered by three methods. (1) Our method; (2) Lv's method; (3) Three-frequency twelve-step phase-shifting method.

Using the absolute phase obtained by the three-frequency twelve-step phase-shifting method as the standard value, we calculate the absolute error and root mean square error between the absolute phase obtained by our method and Lv's method and the standard value.

Fig. 17 shows the absolute error and RMSE. Fig. 17(1) shows our method. Fig. 17(2) shows Lv's method.

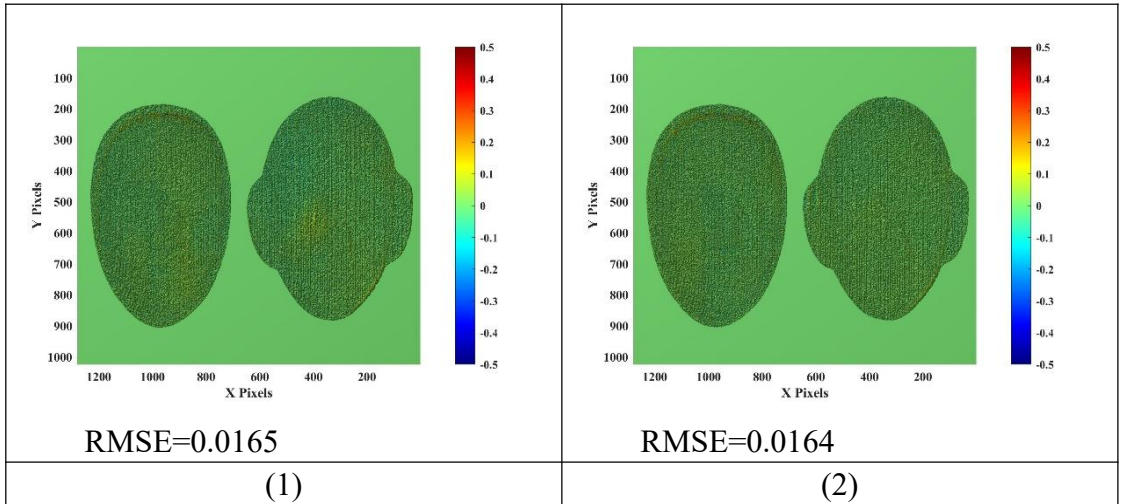

Figure17. The absolute error and RMSE. (1) Our method; (2) Lv's method.

By contrast, obviously, both our methods and Lv's method can retrieve the surface well. It can be seen from the absolute error and RMSE that our method and Lv's method are almost the same in precision.

In the second experiment, the tested objects are one isolated masks and one isolated prism. They are measured by our method, Lv's method and three-frequency twelve-step phase-shifting method respectively. The parameters of the three methods are the same as experiment 1.

Fig. 18 are images taken by our method.

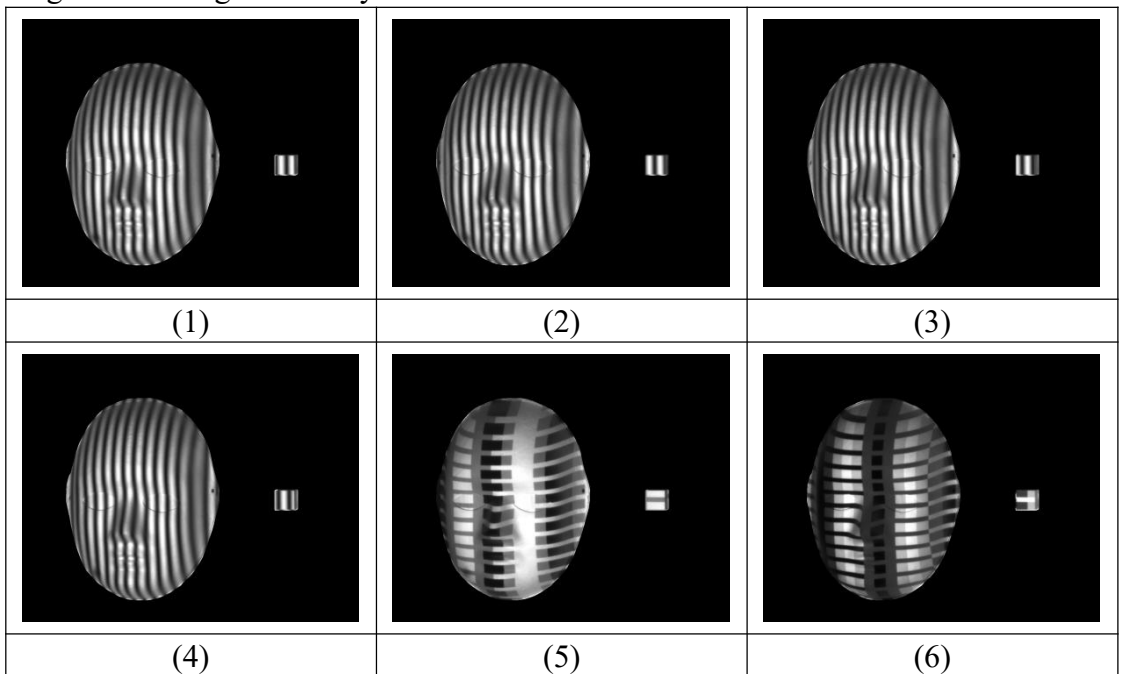

Figure 18. Images taken by our method. (1)-(4) Four-step phase-shifting fringe patterns. (5)-(6) Bidirectional coded stair phase coding patterns.

Fig. 19 shows the absolute phase recovered by our method, Lv's method and three-frequency twelve-step phase-shifting method. Fig. 19(1) shows the absolute phase recovered by our method. Fig. 19(2) shows the absolute phase recovered by

Lv's method. Fig. 19(3) shows the absolute phase recovered by the three-frequency twelve-step phase-shifting method.

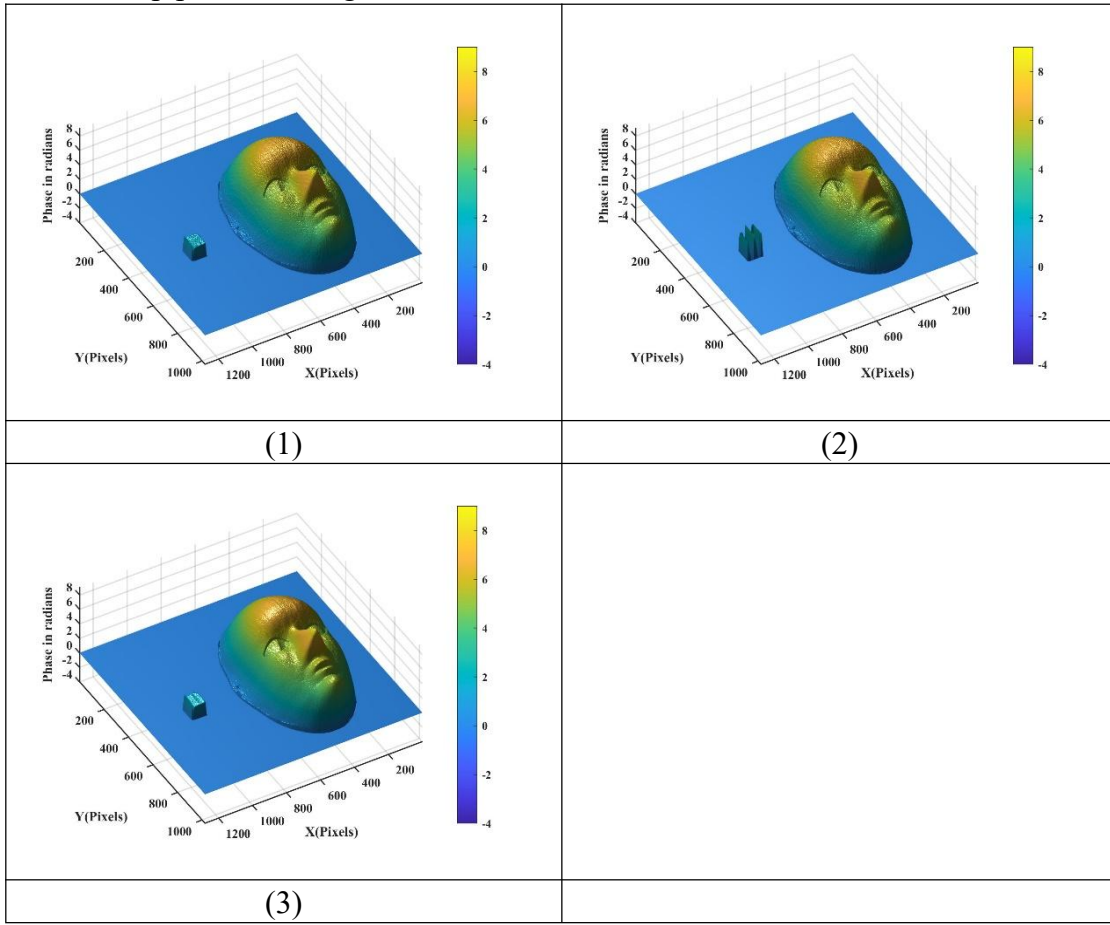

Figure 19. The absolute phase recovered by three methods. (1) Our method; (2) Lv's method; (3) Three-frequency twelve-step phase-shifting method.

It can be seen that due to the few number of fringe on the prism, Lu's method cannot compare the adjacent fringe to obtain the fringe order, and cannot recover the absolute phase of the prism.

Using the absolute phase obtained by the three-frequency twelve-step phase-shifting method as the standard value, we calculate the absolute error and root mean square error between the absolute phase obtained by our method and Lv's method and the standard value.

Fig. 20 shows the absolute error and RMSE.

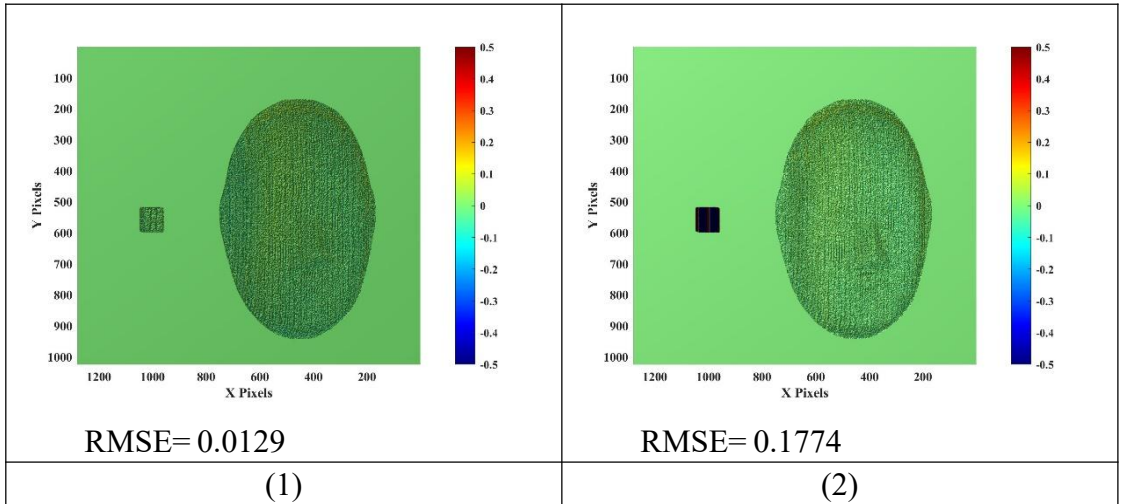

| RMSE= 0.0129 | RMSE= 0.1774 |
| --- | --- |
| (1) | (2) |

Figure 20. The absolute error and RMSE. (1) Our method; (2) Lv's method.

When measuring more than two objects with a large size difference, Lv's method requires at least four adjacent codewords phase information to determine their fringe order. The absolute phase of a small object may not be recovered due to the few number of fringe on it. But our method can recover the absolute phase of all objects.

In the third experiment, the tested objects are two isolated sculptures. They are measured by our method and three-frequency twelve-step phase-shifting method respectively. In our method, the fringe frequency is 1/30 pixel, the number of partitions is 8, the number of stairs in each region is 8, and the total fringe number is 64. The fringe frequency of three-frequency twelve-step phase-shifting method is 1/1920 pixel, 1/240 pixel and 1/30 pixel respectively. Due to the different number of fringes, this experiment was not compared with Lv's method.

Fig. 21 are images taken by our method.

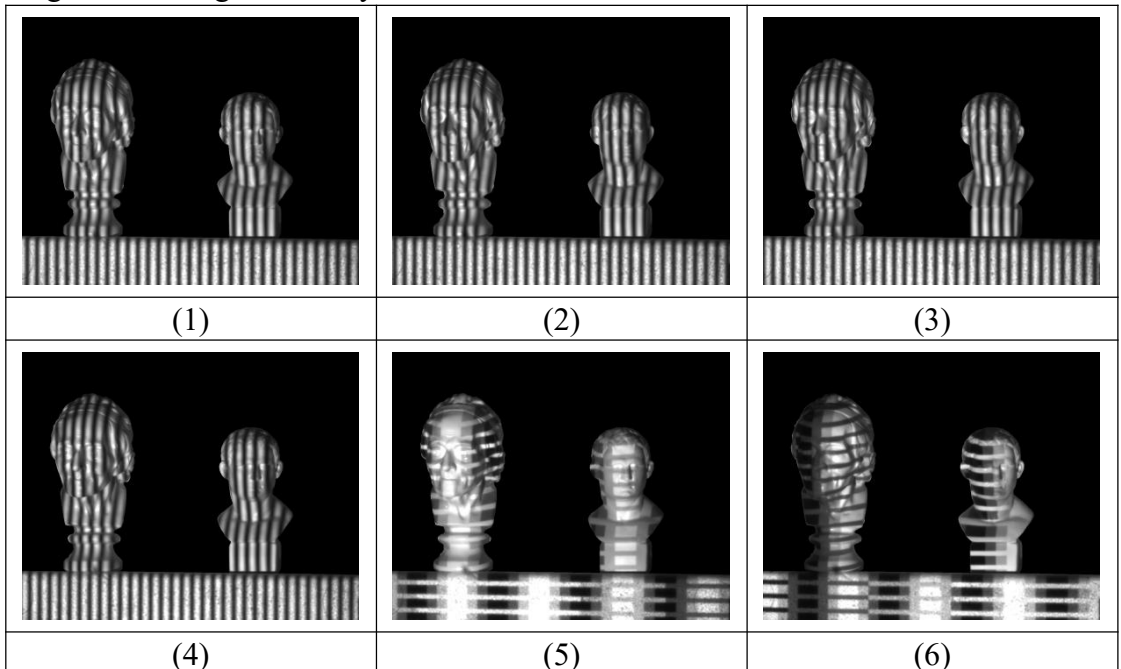

Figure 21. Images taken by our method. (1)-(4) Four-step phase-shifting fringe patterns. (5)-(6) Bidirectional coded stair phase coding patterns.

Fig. 22 shows the absolute phase recovered by the two methods. Fig. 22(1)-(2) shows the absolute phase recovered by our method. Fig. 22 (3)-(4) shows the absolute phase recovered by the three-frequency twelve-step phase-shifting method.

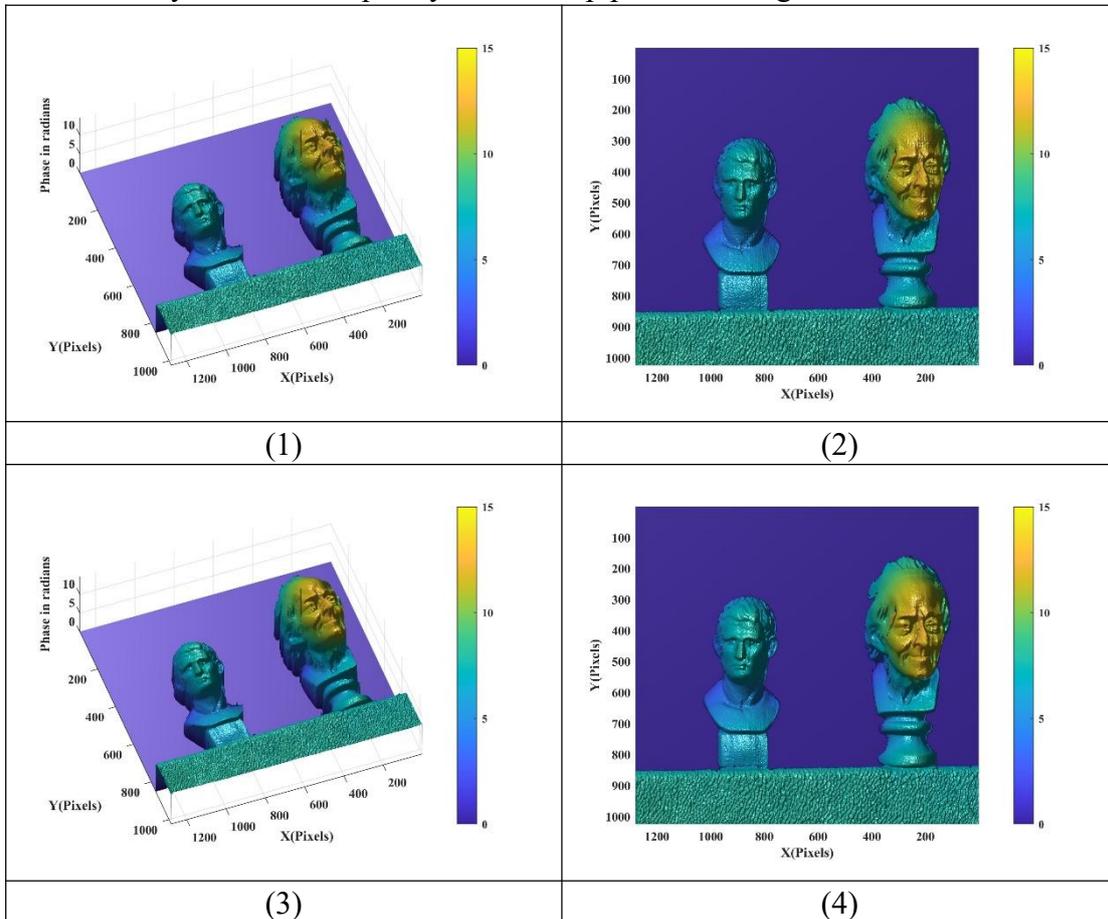

Figure 22. shows the absolute phase recovered by the two methods. (1)-(2) Stereoscopic and flat display of our method; (3)-(4) Stereoscopic and flat display of three-frequency twelve-step phase-shifting method.

Fig. 23 shows the absolute error and RMSE.

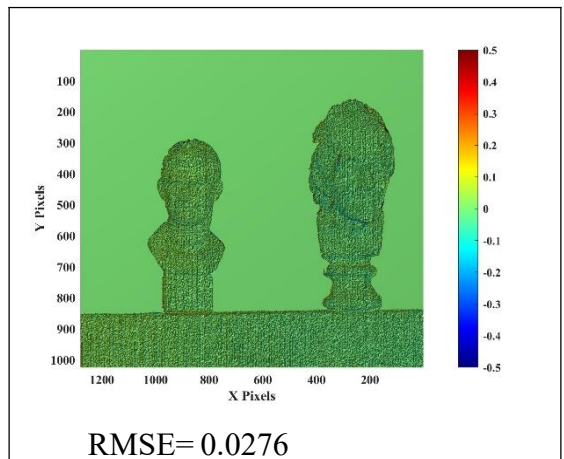

Figure 23. The absolute error and RMSE.

According to absolute error and RMSE, our method is also effective in measuring isolated complex surfaces.

## 5. Conclusion

In this paper, we presented an absolute phase measurement method based on bidirectional stair phase coding. The wrapped phase was calculated by four-step phase-shifting patterns and the fringe order was calculated by two bidirectional stair phase patterns that alternately encodes the local fringe order information and the partition information vertically. The proposed method has the advantages of large number of codewords and low mis-identification rate for fast 3D surface measurement. Compared with Lu's method, the accuracy of our method is almost the same when it is used in relatively simple scenarios, but we can use a set of local fringe order information and partition information in a very small range to determine the fringe order, which has an obvious advantage in measuring more than two isolated objects with large size difference. Simulation and experimental results confirm the effectiveness and superiority of our method.

## ACKNOWLEDGMENT

This work was supported by the National Natural Science Foundation of China (Grant no:11672162). The support is gratefully acknowledged.

## REFERENCES

[1]  Wu Z, Guo W, Zhang Q. Two-frequency phase-shifting method vs. Gray-coded-based method in dynamic fringe projection profilometry: A comparative review[J]. Optics and Lasers in Engineering, 2022, 153: 106995.

[2]  Wu G, Yang T, Liu F, et al. Suppressing motion-induced phase error by using equal-step phase-shifting algorithms in fringe projection profilometry[J]. Optics Express, 2022, 30(11): 17980-17998.

[3]  Pei X, Liu J, Yang Y, et al. Phase-to-Coordinates Calibration for Fringe Projection Profilometry Using Gaussian Process Regression[J]. IEEE Transactions on Instrumentation and Measurement, 2022, 71: 1-12.

[4]  Wang J, Zhang Z, Lu W, et al. High-Accuracy Calibration of High-Speed Fringe Projection Profilometry Using a Checkerboard[J]. IEEE/ASME Transactions on Mechatronics, 2022.

[5]  Li Y, Qian J, Feng S, et al. Composite fringe projection deep learning profilometry for single-shot absolute 3D shape measurement[J]. Optics Express, 2022, 30(3): 3424-3442.

[6]  Ding Y, Zhou C, Qi X, et al. Direct gamma estimation method in Fringe projection profilometry[C]//Eighth Symposium on Novel Photoelectronic Detection Technology and Applications. SPIE, 2022, 12169: 3280-3285.

[7]  Sun J, Zhang Q. A 3D shape measurement method for high-reflective surface based on accurate adaptive fringe projection[J]. Optics and Lasers in Engineering, 2022, 153: 106994.

[8]  Wu H, Cao Y, An H, et al. Ultrafast spatial phase unwrapping algorithm with accurately correcting transient phase error[J]. Optics letters, 2021, 46(24): 6091-6094

[9] Zhang S. Absolute phase retrieval methods for digital fringe projection profilometry: A review[J]. Optics and Lasers in Engineering, 2018, 107: 28-37

[10] He X, Kemao Q. A comparative study on temporal phase unwrapping methods in high-speed fringe projection profilometry[J]. Optics and Lasers in Engineering, 2021, 142: 106613.

[11] Lei Z, Wang C, Zhou C. Multi-frequency inverse-phase fringe projection profilometry for nonlinear phase error compensation[J]. Optics and Lasers in Engineering, 2015, 66: 249-257.

[12] Wang Y, Zhang S. Novel phase-coding method for absolute phase retrieval[J]. Optics letters, 2012, 37(11): 2067-2069

[13] Cai B, Yang Y, Wu J. An improved gray-level coding method for absolute phase measurement based on half-period correction. Optics and Lasers in Engineering,2020;128:106012.

[14] An Yatong, Zhang Song. Three-dimensional absolute shape measurement by combining binary statistical pattern matching with phase-shifting methods. Appl. Opt.,2017;56:5418–26.

[15] Lohry William, Chen Vincent, Zhang Song. Absolute three-dimensional shape measurement using coded fringe patterns without phase unwrapping or projector calibration. Opt. Express 2014;22:1287–301

[16] Zhang B, Lin S, Lin J, et al. Single-shot high-precision 3D reconstruction with color fringe projection profilometry based BP neural network[J]. Optics Communications, 2022, 517: 128323.

[17] Li Y X, Qian J, Feng S, et al. Deep-learning-enabled dual-frequency composite fringe projection profilometry for single-shot absolute 3D shape measurement[J]. Opto-Electron. Adv, 2022, 5: 210021.

[18] Yao P, Gai S, Da F. Toward Real-World Super-Resolution Technique for Fringe Projection Profilometry[J]. IEEE Transactions on Instrumentation and Measurement, 2022, 71: 1-8.

[19] Zhou C, Liu T, Si S, et al. An improved stair phase encoding method for absolute phase retrieval[J]. Optics and Lasers in Engineering, 2015, 66: 269-278.

[20] Zheng D, Da F. Phase coding method for absolute phase retrieval with a large number of codewords. Opt Express 2012;20(22):24139–50

[21] Chen X, Wu J, Fan R, et al. Two-Digit Phase-Coding Strategy for Fringe Projection Profilometry. IEEE Transactions on Instrumentation and Measurement, 70,7001309 (2020)

[22] Y. Xing, C. Quan, and C. J. Tay, "A modified phase-coding method for absolute phase retrieval," Opt. Lasers Eng., 2016,vol. 87, : 97–102

[23] Wang Y, Liu L, Wu J, et al. Enhanced phase-coding method for three-dimensional shape measurement with half-period codeword[J]. Applied optics, 2019, 58(27): 7359-7366

[24] Zheng Y, Jin Y, Duan M, et al. Joint Coding Strategy of the Phase Domain and Intensity Domain for Absolute Phase Retrieval[J]. IEEE Transactions on Instrumentation and Measurement, 2021, 70: 1-8.

[25] Zhang Y, Tong J, Lu L, et al. Fringe Order Correction for Fringe Projection Profilometry Based on Robust Principal Component Analysis[J]. IEEE Access, 2021, 9: 23110-23119

[26] WANG Y, LIU L, WU J, et.al. Dynamic three-dimensional shape measurement with a complementary phase-coding method[J]. Optics and Lasers in Engineering, 2020, 127: 105982

[27] Wu Z, Guo W, Lu L, et al. Generalized phase unwrapping method that avoids jump errors for fringe projection profilometry[J]. Optics Express, 2021, 29(17): 27181-27192

[28] Cai B, Shen Y, Wu J, et al. Optimized phase-coding method for absolute phase retrieval based on K-means algorithm[J]. Journal of Modern Optics, 2021, 68(6): 303-310

[29] Lv S, Jiang M, Su C, et al. Improved unwrapped phase retrieval method of fringe projection profilometry system based on fewer phase-coding patterns[J]. Applied Optics, 2019, 58(32): 8993-9001

[30] Chen X, Wang Y, Wang Y, et al. Quantized phase coding and connected region labeling for absolute phase retrieval[J]. Optics Express, 2016, 24(25):28613

[31] Shiyang T, Yanjun F, Jiannan G, et al. A novel fast 3D measurement method based on phase-coded fringe projection[J]. Optical Review, 2022: 1-10

[32] Cai B, Zhang L, Wu J, et al. Absolute phase measurement with four patterns based on variant shifting phases[J]. Review of Scientific Instruments, 2020, 91(6): 065115.

[33] Li W, Fan N, Wu Y, et al. Fringe-width encoded patterns for 3D surface profilometry[J]. Optics Express, 2021, 29(21): 33210-33224